\documentclass{article}
\usepackage[a4paper, margin=1in]{geometry}
\usepackage{color}
\usepackage{booktabs}
\usepackage{comment}
\usepackage{amsmath}
\usepackage{amssymb}
\usepackage{tabularx}
\usepackage{hyperref}
\usepackage{authblk}
\usepackage{caption}
\usepackage{subcaption}
\usepackage{graphicx}
\usepackage{algpseudocode}
\usepackage[ruled,vlined]{algorithm2e}
\usepackage[skip=0.7em]{parskip}

\hypersetup{
	unicode=false,          
	pdftoolbar=true,        
	pdfmenubar=true,        
	pdffitwindow=false,     
	pdfstartview={FitH},    
	pdftitle={My title},    
	pdfauthor={Author},     
	pdfsubject={Subject},   
	pdfcreator={Creator},   
	pdfproducer={Producer}, 
	pdfkeywords={keywords}, 
	pdfnewwindow=true,      
	colorlinks=true,        
	linkcolor=blue,         
	citecolor=blue,         
	filecolor=magenta,      
	urlcolor=cyan           
}

\title{
    \textbf{Social welfare optimisation in well-mixed and structured populations  }
}
\author[1,2]{Van An Nguyen}
\author[1,2]{Vuong Khang Huynh}
\author[1,2]{Ho Nam Duong}

\author[1,2]{Huu Loi Bui}
\author[1,2]{Hai Anh Ha}
\author[1,2]{Quang Dung Le}
\author[1,2]{Le Quoc Dung Ngo}
\author[1,2]{Tan Dat Nguyen}
\author[1,2]{Ngoc Ngu Nguyen}
\author[1,2]{Hoai Thuong Nguyen}
\author[3]{Zhao Song}

\author[1,2,$\star$]{Le Hong Trang}
\author[3,$\star$]{The Anh Han }

\affil[1]{Faculty of Computer Science and Engineering, Ho Chi Minh City University of
Technology (HCMUT), Vietnam}
\affil[2]{Vietnam National University - Ho Chi Minh City (VNU-HCM), Vietnam}
\affil[3]{School of Computing, Engineering and Digital Technologies, Teesside University, United Kingdom}
\affil[$\star$]{Corresponding authors: Le Hong Trang(Email: lhtrang@hcmut.edu.vn), The Anh Han (Email: t.han@tees.ac.uk)}
\date{ }

\begin{document}

\maketitle

\begin{abstract}

Research on promoting cooperation among autonomous, self--regarding agents has often focused on the bi--objective optimization problem: minimizing the total incentive cost while maximising the frequency of cooperation. However, the optimal value of social welfare under such constraints remains largely unexplored. In this work, we hypothesise that achieving maximal social welfare is not guaranteed by the minimal incentive cost required to drive agents to a desired cooperative state. To address this gap, we adopt to a single--objective approach focused on maximising social welfare, building upon foundational evolutionary game theory models that examined cost efficiency in finite populations, in both well-mixed and structured population settings. Our analytical model and agent--based simulations show how different interference strategies, including rewarding local versus global behavioural patterns, affect social welfare and dynamics of cooperation. Our results reveal a significant gap in the per--individual incentive cost between optimising for pure cost efficiency or cooperation frequency and optimising for maximal social welfare.
Overall, our findings indicate  that incentive design, policy, and benchmarking in  multi‑agent systems and human societies should prioritise welfare‑centric objectives over proxy targets of cost or  cooperation frequency.
\end{abstract}

\section{Introduction}

\begin{figure*}[ht]
    \centering
    \includegraphics[width=.7\linewidth]{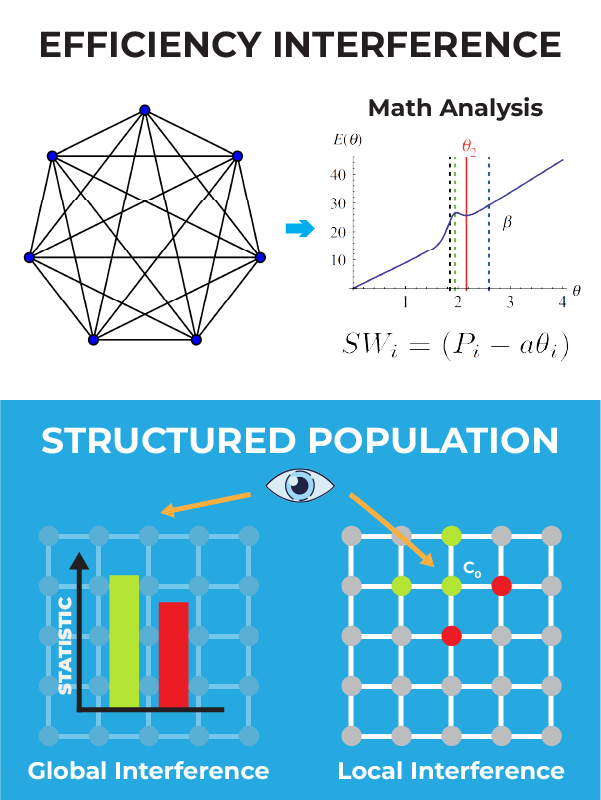}
    \caption{
    Overall approach. By exploring the use of Social Welfare as a metric of
    optimisation for the evolutionary model of cooperation under the effect of
    institutional incentives, this work aims to answer the differentiation
    between cost optimisation and social welfare optimisation both theoretically
    and empirically.
    }
    \label{fig:overall-approach}
\end{figure*}

The evolution of cooperation has long been a central puzzle in evolutionary biology, social sciences and multi--agent systems  \cite{sigmund2010calculus,perc2017statistical,han2022emergent,paiva2018engineering,tuyls2007evolutionary}. 
While classical evolutionary theory emphasizes the survival of the fittest--- which in many strategic 
settings (e.g., the Prisoner's Dilemma) often corresponds to more selfish behaviour---cooperative 
behaviour is nevertheless pervasive among both humans and animals. This apparent contradiction has 
motivated extensive research into the mechanisms that enable and explain how cooperation emerges and persists in social dilemmas \cite{nowak2006five}.

To address this puzzle of cooperation \cite{perc2017statistical,rand2011evolution}, numerous mechanisms have been proposed to account for the emergence of cooperation, including kin and group selection \cite{hamilton1964genetical,traulsen2006evolution}, direct and indirect reciprocity \cite{nowak:2005:nature,axelrod1981evolution}, reward and punishment (incentives) \cite{fehr2000cooperation,chen2015first}  and structured populations \cite{szabo2007evolutionary}.
A particularly important mechanism examined in recent years involves an external institution 
that seeks to steer the population towards greater cooperation by selectively investing in individuals \cite{chen2015first,duong2023cost,wang2022incentive,sigmund2001reward}. 
For example, such an institution may provide rewards to cooperators, either at the global population 
level or in a more localized, neighbourhood--based manner. 
However, existing research often overlooks \textit{social} welfare and fails to address the delicate balance between fostering cooperation and optimizing social welfare.

In well--mixed populations, Han and Tran--Thanh (2018) \cite{han2018interference} showed that a decision--maker can conditionally 
reward cooperators based on population composition to guarantee a desired cooperation level, while 
minimizing interference cost. Their framework formulates a bi--objective optimisation problem: maximize 
cooperation frequency while minimizing institutional investment. 
Later, Duong and Han (2021) \cite{duonghan2021cost} provided a rigorous stochastic analysis of institutional incentives, 
characterizing optimal reward/punishment schemes under different selection intensities and identifying 
sharp phase transition phenomena in cost efficiency.

However, real populations are seldom well--mixed. Interaction patterns are shaped by spatial or network 
structures, which can fundamentally alter evolutionary outcomes \cite{szabo2007evolutionary,perc2017statistical,nowak1992evolutionary}. Han et al.\ (2018) \cite{han2018fostering} examined external 
interference in structured populations using agent--based simulations on square lattices (which was then extended to other networks and game--theoretic interactions \cite{cimpeanu2021cost,cimpeanu2019exogenous}). Their results indicate 
that local interference strategies, which monitor neighbourhood--level information, can be 
significantly more cost--efficient than global ones, underscoring the importance of spatial 
heterogeneity when designing incentive mechanisms.

Despite these advances,  prior works mainly  focus on either (i) maximising cooperation prevalence or 
(ii) minimizing the institutional cost \cite{duonghan2021cost,wang2019exploring}. What is missing is a direct optimisation of social welfare, 
 defined as:
\[
\text{Social Welfare} = \text{(Total payoff of the population)} - \text{(External cost EC)}.
\]

From a societal or institutional perspective, social welfare is ultimately the most meaningful 
objective: cooperation is valuable insofar as it generates net benefits relative to the cost of 
enforcing it \cite{karsu2015inequity,nyman2006efficiency,HanInterface2024Welfare,HAN2026_social_welfare}.
This paper introduces social welfare maximization into the study of 
cost--efficient external interference. 
We address the following research questions:
    \begin{itemize}
        \item \textbf{RQ1}: How does optimizing social welfare change the optimal interference 
        strategies in well--mixed populations? Does welfare maximization require less, more, or 
        differently patterned investment compared to minimizing cost alone?
        \item \textbf{RQ2}: In structured populations, do previously identified cost--efficient local 
        strategies remain superior under the welfare objective? Does spatial structure amplify or 
        diminish the welfare benefits of conditional interference?
    \end{itemize}

Our key contributions are summarized as follows:
\begin{itemize}
    \item We incorporate social welfare into the analytical framework of institutional incentives in 
    well--mixed populations, extending the interference scheme of Han \& Tran--Thanh (2018) \cite{han2018interference} and the 
    cost--efficiency analysis of Duong \& Han (2021) \cite{duonghan2021cost}. This allows us to examine how welfare maximization 
    reshapes optimal intervention strategies.

    \item We evaluate interference in spatially structured populations on grids using agent--based 
    simulations, extending Han et al.\ (2018). We compare global and local interference strategies not 
    only in terms of cooperation frequency or cost efficiency, but through the unified lens of social 
    welfare.
\end{itemize}

Our overall approach can be summarized in Figure \ref{fig:overall-approach}. By introducing social welfare into evolutionary models of institutional incentives, this work 
offers a more realistic and policy--relevant framework for understanding how cooperation can be 
engineered in biological, social, and artificial systems. 
Our unified perspective bridges analytical well--mixed models and simulation--based structured models, 
shedding new light on the design of welfare--maximizing intervention schemes.

\section{Literature review}

The evolution of cooperation has been a central focus in evolutionary game theory, where social dilemmas such as the Prisoner’s Dilemma and Public Goods Game illustrate the tension between individual incentives and collective benefits~\cite{archetti2012game,axelrod1981evolution,doebeli2005models}. In one--shot interactions, defection strictly dominates cooperation, yet cooperative behaviour is widely observed across biological, social, and artificial systems~\cite{fehr2004social}. This apparent contradiction has motivated extensive research on mechanisms that enable cooperation to emerge and persist, including reciprocity~\cite{nowak2006five,nowak2005evolution}, spatial structure~\cite{szabo2007evolutionary}, kin selection~\cite{hamilton1964genetical}, and institutional regulation~\cite{sasaki2012take}.

A prominent line of work examines external institutional influence, where a central authority invests resources to steer population behaviour~\cite{duong2023cost,sigmund2001reward,cimpeanu2021cost}. Early models typically assumed well--mixed populations and studied how rewards, punishments, or other incentive schemes affect the stability of cooperation. Han and Tran--Thanh (2018) introduced a conditional investment mechanism in which an external controller (EC) allocates rewards depending on the current population state~\cite{han2018interference}. Their analysis showed that the EC can guarantee a desired cooperation level while minimizing the total interference cost, formally framing the problem as a bi--objective optimisation between maximizing cooperation and minimizing investment. Building on this, Duong and Han (2021) provided a more comprehensive stochastic treatment, deriving closed--form results for cost efficiency under various selection intensities and revealing non-trivial threshold behaviours in the optimal interference levels~\cite{duonghan2021cost}.

Beyond well--mixed populations, spatial and network structures introduce additional complexity~\cite{santos2006cooperation,perc2013evolutionary,nowak1992evolutionary}. Real systems often involve locality constraints---agents interact more frequently with a subset of neighbours rather than the entire population. Han et al. (2018) extended institutional incentive models to structured populations using agent--based simulations on grids~\cite{han2018fostering}. They demonstrated that local interference strategies, which condition investment on neighbourhood--level information, can outperform global strategies in terms of both cooperation levels and cost efficiency. These results highlight the importance of spatial heterogeneity in designing external incentive mechanisms.

Despite these important advances, most existing models evaluate institutional performance using only two criteria: maximizing cooperation or minimizing the external cost~\cite{gross2025hidden}. While informative, these objectives do not fully capture the system--level impact of external intervention. In scenarios where an institution must balance the benefits generated by cooperation against the resources required to sustain it, social welfare---defined as the total payoff of the population minus the external investment---provides a more holistic measure~\cite{han2025cooperation,duong2024evolutionary,kaneko1979nash}. Although some preliminary studies have mentioned welfare concepts, a systematic analysis of optimal interference strategies under a welfare--maximization framework is still lacking in both analytical well--mixed models and simulation--based structured models.

This gap raises several important questions. In well--mixed populations, does the interference scheme that minimises cost or maximizes cooperation also maximize social welfare? Similarly, in structured populations, do local interference strategies---previously shown to be cost--efficient---remain optimal when evaluated with respect to social welfare? Furthermore, do spatial constraints amplify or diminish the welfare benefits of conditional interference?

In summary, while prior works have laid a solid foundation for understanding institutional incentives in evolutionary dynamics, a unified social welfare--based analysis across both well--mixed and structured populations is still missing. Addressing this gap is essential not only for theoretical completeness but also for practical relevance, especially in contexts such as public policy, distributed AI systems, and resource allocation, where maximizing net societal benefit is a central goal.

\section{Models and Methods}
This section reviews the original Evolutional Game Theory model for previous optimization problems in \cite{duonghan2021cost}, and derives the  social welfare function in a well--mixed population setting. Furthermore, it  describes  the agent--based simulation setup extending from \cite{han2018fostering}, including the simulation process, the formulation of social welfare, and the different intervention strategies that this work will study.

\subsection{Problem setup}

We consider a well--mixed, finite population of $N$ players, who interact with each other using cooperation dilemmas, such as the Donation Game or Public Goods Game. Let $\Pi_C(i) \text{ and } \Pi_D(i)$ be the expected payoffs of a cooperative player (C--player) and a defective player (D--player) in state $S_i$ of the population, respectively. $S_i$ represents a state where the population has $i$ C--players. 

\subsubsection*{Donation Game (DG)}

\noindent The Donation Game is a special case of the Prisoners' Dilemma, where cooperation corresponds to providing the co--player with a benefit $b$ at a personal cost $c$, with $b>c$, while defection yields no benefit and incurs no cost. The payoff matrix of the game (for the row player) is given by
\[
 \bordermatrix{~ & C & D\cr
                  C & b-c & -c \cr
                  D & b & 0  \cr
                 }.
\]

\noindent Let $\pi_{X,Y}$ denote the payoff of a player using strategy $X \in \{C,D\}$ when interacting with a player using strategy $Y \in \{C,D\}$. In a well--mixed population of size $N$, at state $S_i$ where there are $i$ cooperators, the expected payoffs of a C--player and a D--player are given by
\begin{equation*}
\begin{aligned}
\Pi_C(i) &= \frac{(i-1)\pi_{C,C} + (N-i)\pi_{C,D}}{N-1}
         = \frac{(i-1)(b-c) + (N-i)(-c)}{N-1}, \\
\Pi_D(i) &= \frac{i\pi_{D,C} + (N-i-1)\pi_{D,D}}{N-1}
         = \frac{ib}{N-1}.
\end{aligned}
\end{equation*}

\noindent Therefore, the payoff difference between cooperation and defection is
\[
\delta = \Pi_C(i) - \Pi_D(i) = -\Big(c + \frac{b}{N-1}\Big),
\]
which is negative and independent of the population state $S_i$, in accordance with the general assumption introduced earlier.

\subsubsection*{Public Goods Game (PGG)}

\noindent In the Public Goods Game, individuals interact in groups of size $n$. Each player can either cooperate by contributing an amount $c>0$ to a common pool, or defect by contributing nothing. The total contribution within a group is multiplied by an enhancement factor $r$, with $1<r<n$, and the resulting amount is equally shared among all group members, independently of their strategies. Since defectors benefit from the public good without paying the cost, the game constitutes a social dilemma.

\noindent In a well--mixed population of size $N$, at state $S_i$ where there are $i$ cooperators, groups are formed by multivariate hypergeometric sampling. Hence, the expected payoffs of a C--player and a D--player are given by
\begin{equation*}
\begin{aligned}
\Pi_C(i)
&= \sum^{n-1}_{k=0}
\frac{\dbinom{i-1}{k}\dbinom{N-i}{\,n-1-k\,}}{\dbinom{N-1}{\,n-1\,}}
\left(\frac{(k+1)rc}{n} - c \right)
= \frac{rc}{n}\left(1 + (i-1)\frac{n-1}{N-1}\right) - c, \\[6pt]
\Pi_D(i)
&= \sum^{n-1}_{k=0}
\frac{\dbinom{i}{k}\dbinom{N-1-i}{\,n-1-k\,}}{\dbinom{N-1}{\,n-1\,}}
\frac{krc}{n}
= \frac{rc(n-1)}{n(N-1)}\, i .
\end{aligned}
\end{equation*}

\noindent Therefore, the payoff difference between cooperation and defection is
\[
\delta = \Pi_C(i) - \Pi_D(i)
= -c \left(1 - \frac{r(N-n)}{n(N-1)} \right),
\]
which is negative and independent of the population state $S_i$.

\subsubsection*{Cost of institutional reward and punishment}

To reward a cooperator (respectively, punish a defector), the institution has to spend an amount $\theta/a$ (respectively, $\theta/\hat{a}$), such that the payoff of the targeted individual increases (or decreases) by $\theta$, where $a,\hat{a}>0$ denote the efficiency ratios of reward and punishment, respectively.

In an institutional enforcement setting, we assume that the institution has full information about the population composition at the time of decision--making. Namely, in the well--mixed population of size $N$, the number $i$ of cooperators in state $S_i$ is known. If both reward and punishment are feasible (mixed incentives), the institution minimises its instantaneous cost by choosing the cheaper option between rewarding all cooperators and punishing all defectors, that is
\[
\min\Big(\frac{i}{a},\frac{N-i}{\hat{a}}\Big)\theta.
\]
The central question is thus: what is the minimal incentive intensity $\theta$ that ensures a prescribed long--run level of cooperation while minimizing the total institutional cost?

\subsubsection*{Expected cost of institutional incentives}

We consider a finite, well--mixed population evolving according to the Fermi strategy update rule~\cite{sigmund2010social}. A player $X$ with fitness $f_X$ adopts the strategy of another player $Y$ with fitness $f_Y$ with probability
\[
P_{X,Y}=\left(1+e^{-\beta(f_Y-f_X)}\right)^{-1},
\]
where $\beta>0$ denotes the selection intensity.

The population dynamics are modelled as an absorbing Markov chain over the state space $\{S_0,S_1,\dots,S_N\}$, where $S_i$ represents the state with $i$ cooperators. The homogeneous states $S_0$ and $S_N$ are absorbing, while $S_1,\dots,S_{N-1}$ are transient. Let $U=\{u_{ij}\}_{i,j=1}^{N-1}$ be the transition matrix among transient states. For $1\le i\le N-1$, the transition probabilities are
\begin{equation}
\label{eq: transition probabilities}
\begin{aligned}
u_{i,i\pm k} &= 0, \qquad k\ge 2, \\
u_{i,i+1} &= \frac{N-i}{N}\frac{i}{N}
\left(1+e^{-\beta[\Pi_C(i)-\Pi_D(i)+\theta]}\right)^{-1}, \\
u_{i,i-1} &= \frac{N-i}{N}\frac{i}{N}
\left(1+e^{\beta[\Pi_C(i)-\Pi_D(i)+\theta]}\right)^{-1}, \\
u_{i,i} &= 1-u_{i,i+1}-u_{i,i-1}.
\end{aligned}
\end{equation}

Let $\mathcal{N}=(I-U)^{-1}=(n_{ik})_{i,k=1}^{N-1}$ denote the fundamental matrix of this chain. The entry $n_{ik}$ gives the expected number of visits to state $S_k$ when starting from state $S_i$. Since mutants can appear with equal probability in $S_0$ and $S_N$, the expected number of visits to $S_j$ is
\[
\frac{1}{2}(n_{1j}+n_{N-1,j}).
\]

The instantaneous cost of providing incentives in state $S_j$ is therefore
\[
\theta_j=\min\Big(\frac{j}{a},\frac{N-j}{\hat{a}}\Big)\theta.
\]
Hence, the expected total cost of mixed incentives is
\begin{equation}
\label{eq:total_investment}
E_{\text{mix}}(\theta)
= \frac{\theta}{2}\sum_{j=1}^{N-1}(n_{1j}+n_{N-1,j})
\min\Big(\frac{j}{a},\frac{N-j}{\hat{a}}\Big).
\end{equation}

For comparison, the expected costs of using only reward and only punishment are given by 
\begin{equation}
\label{eq: Er and Ep}
E_r(\theta)=\frac{\theta}{2}\sum_{j=1}^{N-1}(n_{1j}+n_{N-1,j})\frac{j}{a},
\qquad
E_p(\theta)=\frac{\theta}{2}\sum_{j=1}^{N-1}(n_{1j}+n_{N-1,j})\frac{N-j}{\hat{a}}.
\end{equation}

We note that, unlike the separate reward and punishment schemes, the efficiency ratios $a$ and $b$ directly affect the structure of the mixed incentive cost $E_{\text{mix}}(\theta)$ through their interaction in the minimum operator, making the optimisation problem  more complex.

\subsubsection*{Cooperation frequency}

Since the population consists of two strategies, the fixation probabilities of a single cooperator in a population of defectors and vice versa are given by 
\begin{equation*}
\begin{aligned}
\rho_{D,C}
&=\left(1+\sum_{i=1}^{N-1}
\prod_{k=1}^{i}
\frac{1+e^{\beta[\Pi_C(k)-\Pi_D(k)+\theta]}}
{1+e^{-\beta[\Pi_C(k)-\Pi_D(k)+\theta]}}
\right)^{-1}, \\
\rho_{C,D}
&=\left(1+\sum_{i=1}^{N-1}
\prod_{k=1}^{i}
\frac{1+e^{\beta[\Pi_D(k)-\Pi_C(k)-\theta]}}
{1+e^{-\beta[\Pi_D(k)-\Pi_C(k)-\theta]}}
\right)^{-1}.
\end{aligned}
\end{equation*}

The stationary frequency of cooperation is then
\[
\frac{\rho_{D,C}}{\rho_{D,C}+\rho_{C,D}}.
\]
Maximising this frequency is equivalent to maximising
\begin{equation}
\label{eq:max}
\max_{\theta}\left(\frac{\rho_{D,C}}{\rho_{C,D}}\right).
\end{equation}

This ratio simplifies as 
\begin{eqnarray}
\nonumber
\frac{\rho_{D,C}}{\rho_{C,D}}
&=&\prod_{k=1}^{N-1}
\frac{u_{k,k-1}}{u_{k,k+1}}
=\prod_{k=1}^{N-1}
\frac{1+e^{\beta[\Pi_C(k)-\Pi_D(k)+\theta]}}
{1+e^{-\beta[\Pi_C(k)-\Pi_D(k)+\theta]}} \\
\nonumber
&=& e^{\beta\sum_{k=1}^{N-1}(\Pi_C(k)-\Pi_D(k)+\theta)} \\
\label{eq:max_Q_prime}
&=& e^{\beta(N-1)(\delta+\theta)}
\end{eqnarray}

Given a desired cooperation level $\omega\in[0,1]$, i.e.
\[
\frac{\rho_{D,C}}{\rho_{D,C}+\rho_{C,D}}\ge \omega,
\]
it follows from \eqref{eq:max_Q_prime} that
\begin{equation}
\label{eq:omega_fraction}
\theta \ge \theta_0(\omega)
= \frac{1}{(N-1)\beta}
\log\!\left(\frac{\omega}{1-\omega}\right)-\delta.
\end{equation}
Thus, whenever $\theta\ge \theta_0(\omega)$, the expected fraction of cooperation is at least $\omega$.

\subsubsection*{Cost minimisation problem}

Combining the cooperation constraint with the cost function, the institutional optimisation problem is
\begin{equation} \label{eq:minprob}
\min_{\theta\ge \theta_0(\omega)} E_{\text{mix}}(\theta)
\end{equation}

\subsection{Analytical calculation}
We formally define here the population social welfare. We define $a \in [0,+\infty)$ as the efficiency of the reward mechanism. Specifically, the institution pays a cost $\theta$ to reward a C player a payoff of $a\theta$. A reward mechanism is considered cost-efficient for the institution  if $a \ge 1$. 
We  analyse the mathematical behaviours of the population social welfare  in three cases: $a = 1 \text{, } a < 1 \text{ and } a > 1$. 

We first derive an expression for the expected total social welfare across population states, $SW(\theta)$, under the influence of a  institutional reward $\theta$ for cooperative players.
The social welfare in state $S_i$ (where $i$ is the number of cooperators) is the total payoff received by all players ($P_i$) minus the total institutional cost paid ($\theta_i$). 

The social welfare in state $S_i$ is the total payoff received by all players ($P_i$) minus the total institutional incentives paid ($\theta_i$). We assume the reward is funded from an external budget, meaning the institution's spending does not reduce the population's baseline payoff. The variables are defined as: $\Pi_D(i)=i\Delta$ (Defector payoff structure) and $\delta = \Pi_C(i) - \Pi_D(i)$ (Payoff difference).

The total payoff received is:
\[
P_i = i\big[\Pi_C(i) + a\theta\big] + (N - i)\,\Pi_D(i)
\]
Given the total institutional cost is $\theta_i=i\theta$, the aggregate social welfare in state $S_i$ is:
\begin{align*}
SW_i(\theta)&= P_i - \theta_i \notag \\
 &= i\big[\Pi_C(i) + a\theta\big] + (N - i)\,\Pi_D(i) - i\theta \notag \\
 &= i\Pi_C(i) + (N - i)\Pi_D(i) + i\theta(a-1) \notag \\
 &= i\big(\Pi_C(i) - \Pi_D(i)\big) + N\Pi_D(i) + i(a-1)\theta \notag \\
 &= i\delta + N(i\Delta) + i(a-1)\theta \notag \\
 &=i\big(\delta + N\Delta + (a-1)\theta \big)
\end{align*}

To evaluate the expected population-level social welfare, we aggregate $SW_i(\theta)$ over all transient states $S_i$ ($i = 1$ to $N-1$), weighted by the expected visitation rates ($\frac{n_{1,i} + n_{N-1,i}}{2}$) derived from the Fundamental Matrix ($\mathcal{N}$) defined in \cite{duonghan2021cost}. 
\begin{align}
\nonumber
SW(\theta)=\sum_i SW_i(\theta) 
&= \sum_{i}i\big(\delta + N\Delta+(a-1)\theta \big) 
\frac{n_{1,i} + n_{N-1,i}}{2} \\ 
\nonumber
&= (\delta + N\Delta + (a-1)\theta) 
\sum_{i}i\frac{n_{1,i} + n_{N-1,i}}{2} \\
&= \frac{N^2}{2}\frac{f(x)}{g(x)} \big(\delta + N\Delta + (a-1)\theta\big) 
 \label{eq:SW-theta-final}
\end{align}

With $x=\beta(\theta + \delta)$, $f(x)$ and $g(x)$ are defined as in the electronic supplementary material of \cite{duonghan2021cost} 
\begin{align}
f(x) &= (1 + e^x)\left[ \left(1 + e^x + \cdots + e^{(N-2)x}\right) H_N 
+ e^{(N-1)x} \sum_{j=1}^{N-1} \frac{e^{-jx}}{j} \right] \label{fx}, \\
g(x) &= 1 + e^x + \cdots + e^{(N-1)x} \label{gx}
\end{align}
Note that the summation term $\sum_{i} i \frac{n_{1,i} + n_{N-1,i}}{2}$ is directly related to the expected total cost of institutional reward $E_r(\theta)$ derived in \cite{duonghan2021cost}. 
We use the identity:\[\sum_{i} i \frac{n_{1,i} + n_{N-1,i}}{2} = \frac{E_r(\theta)}{\theta}=\frac{1}{\theta}\left[ \frac{N^2\theta}{2}\frac{f(x)}{g(x)}\right]=\frac{N^2}{2}\frac{f(x)}{g(x)}\]
The analysis of the Expected Total Social Welfare ($SW(\theta)$) depends fundamentally on whether the institutional incentive is a net value creation or a zero--sum transfer. The function $SW(\theta)$ is governed by the efficiency parameter $a$, specifically through the term $(a-1)\theta$. We therefore proceed by analysing two core scenarios: the zero--net--gain case ($a=1$), where the incentive mechanism is a simple transfer and does not directly depend on $\theta$ for its value, and the non--zero--sum case ($a \ne 1$), where the value added or destroyed by the incentive system becomes an explicit function of $\theta$. The final results in \ref{analytical-result} are acquired based on established results in \cite{duonghan2021cost}. 

\subsection{Agent--based simulation}

\subsubsection{Prisoner’s Dilemma on Square Lattice Networks} \label{agent-simulation-setup}

Beyond the well-mixed population assumption above, we distribute agents on a square lattice, a network structure widely used in evolutionary game studies \cite{szabo2007evolutionary,han2018fostering}.
Specifically, we use a network with size $Z = L \times L$. At the beginning, each agent is assigned either as a cooperator (C) or defector (D) with the same probability. We use the one--shot Prisoner’s Dilemma game to model the interaction between agents, where R is the reward for the cooperation (penalty P) and the denial in cooperation gives the defector the temptation T and the fool payoff S. We adopt the scaled payoff matrix of the PD \cite{han2018fostering}: $T=b, R=1, P=S=0$ (with $1 < b \leq 2$).

At each time step or generation, each agent plays the PD  with its four neighbours and decides whether to change the strategy or not based on the neighbour's highest scores. We follow the deterministic, standard evolutionary process in order to capture the cost of different interference strategies for the social welfare efficiency analysis. In addition, we also conduct experiments on a stochastic update rule to prove the correctness of hypothesis. Instead of choosing the highest scores among neighbours, agent A with score $f_A$ has a Fermi probability that it copies the strategy from agent B (randomly chosen among neighbours) with score $f_B$ with the amplitude of noise $K$ ~\cite{traulsen2006stochastic,szabo2007evolutionary}: $(1+e^{(f_A - f_B)/K})^{-1}$.

We run the evolutionary process until it reaches either a steady state or a repeating cycle. To ensure a consistent and fair comparison, each simulation is executed for 50 generations. To further enhance accuracy, the final results for each parameter setting are computed by averaging 10 independent runs. We do not consider mutations or explorations in this work, for a convenient comparison with  \cite{han2018fostering}.

\subsubsection{Social Welfare Efficiency}

As noted above, we hypothesise that minimising interference costs while increasing cooperation does not necessarily maximise population social welfare. We demonstrate this in the present study.

Investors pay a cost $\theta > 0$ for a cooperator (to the external decision--maker/investor). In particular, we investigate whether global interference strategies (where investments are triggered based on population  level information) or their local counterparts (where investments are based on local neighbourhood information) lead to successful behaviour with better social welfare efficiency. To do so, we analyse  two main classes of interference strategies  \cite{han2018fostering} to for their capacity to optimise social welfare: i) global
(population--composition--based -- POP) and ii) local (neighborhood--based -- NEB). 

In the POP strategy, the decision to interfere (i.e. to invest on all cooperators in the population)
is based on the current composition of the population. Specifically, they invest when the number of
cooperators in the population is below a certain threshold, $p_C$ for $1 \leq p_C \leq Z$. The value
$p_C$ describes how widespread defection strategy should be to trigger the support of cooperators’
survival against defectors.

On the other hand, the decision in NEB to invest in a given cooperator is based on the
cooperativeness level in that cooperator’s neighbourhood. In more detail, the decision--maker invests
in a cooperator when the number of its cooperative neighbours is below a certain threshold,
$n_C$, for $0 \leq n_C \leq 4$.

\section{Results}

The first two parts of this section summarise the results of the mathematical analysis of the objective function $SW$ with respect to both $\theta$ and $\beta$, including its variability and limiting behaviours, as well as an approximation algorithm to compute the optimal solution and simulations for various cases of the Donation Game. The latter part of this section discusses the results of the agent--based simulation on a square lattice network, detailing the experimental configurations for both global and local intervention schemes, and utilises figures from different perspectives to interpret the results.

\subsection{Analytical results for well--mixed population}
\label{analytical-result}
We present the core analytical findings regarding the maximization of the Expected Total Social Welfare ($SW$) with respect to the incentive $\theta$. The analysis is organized into two main scenarios: the zero-sum transfer ($a=1$) and the non-zero-sum transfer ($a \ne 1$). Furthermore, this section examines the profound influence of selection intensity ($\beta$) on the objective's behaviour and proposes a numerical algorithm to approximate the optimal incentive for the case of inefficient transfer ($a < 1$).

\subsubsection{Case 1: Zero Sum Transfer (\texorpdfstring{$a=1$}{a=1})}
Under the condition $a=1$, the term $(a-1)\theta$ vanishes, and the Expected Total Social Welfare ($SW(\theta)$) simplifies to:
\[SW(\theta)_{a=1} = K \cdot \Psi(x)\]
where $K = \frac{N^2}{2} (\delta + N\Delta)$ is a positive constant, and $\Psi(x)$ is the Dynamics Factor defined as the ratio $\dfrac{f(x)}{g(x)}$.\\
The problem of maximizing $SW(\theta)_{a=1}$ reduces to finding the optimal evolutionary state $x$ that maximizes the Dynamics Factor $\Psi(x)$. The analysis of the derivative of $\Psi(x)$ w.r.t $\theta$ confirms the existence of a unique maximum:
\[\frac{d\Psi}{d\theta} = -\beta \frac{e^x P(e^x)}{g^2(x)}\]
This derivative was proven to have a unique positive root, $x_0 = \log(u_0) > 0$, where $u_0 > 1$ is the unique positive root of the polynomial $P(u)$. Consequently, $\Psi(x)$ is increasing on $(-\infty, x_0]$ and decreasing on $[x_0, +\infty)$, guaranteeing a global maximum at $x_0$. However, finding the analytical value for $x_0$ is computationally expensive due to the high degree of the polynomial $P(u)$ (defined in supplementary material for \cite{duonghan2021cost}).\\
The maximum overall social welfare is achieved when the incentive $\theta$ forces the system into the state defined by $x_0$. Since $x = \beta(\theta + \delta)$, the Optimal Social Welfare Incentive ($\theta_0^{SW(\theta)}$) is derived as:\[\theta_0^{SW(\theta)}=\dfrac{x_0}{\beta}-\delta\]

\subsubsection{Case 2: Non--Zero Sum Transfer(\texorpdfstring{$a \ne 1$}{a!=1})}
To determine the optimal incentive $\theta$ that maximizes social welfare, we analyze the derivative of $SW(\theta)$ with respect to $\theta$. Substituting the expressions for $f(x)$, $g(x)$, and their derivatives, the derivative $SW(\theta)'(\theta)$ can be factorized into a structured form involving the auxiliary function $F(u)$ (defined in the supplementary material for \cite{duonghan2021cost}): 
\begin{equation} 
\frac{dSW(\theta)}{d\theta} = \frac{N^2 u P(u)}{2 g^2(x)} (a-1) \left[ F(u) + \beta \mathcal{K} \right] 
\end{equation}
where $\mathcal{K}=\dfrac{(a-2)\delta - N\Delta}{a-1}$. We examine the behaviour of $F(u)-\beta \mathcal{K}$ on $(u_0,+\infty)$ ($u_0$ is the unique positive root of $P(u)$), such that $P(u)>0$.\\
\begin{enumerate}
    \item $\mathcal{K} \ge 0$:
    Since $F(u)$ has a unique minimiser $u^*$ on $(u_0,+\infty)$, we have: $F(u)-\beta \mathcal{K} \ge F^* > 0$ ($F^*=F(u^*)$). 
    \begin{itemize}
        \item If $a < 1$: $\dfrac{dSW(\theta)}{d\theta}<0,\forall u\in (u_0, +\infty)$. Therefore:
        \[
        \max_{\theta \ge \theta_0}SW=SW(\theta_0)
        \]
        where $\theta^0=logu_0-\beta \delta$.
        \item If $a > 1$: $\dfrac{dSW(\theta)}{d\theta}>0,\forall u\in (u_0, +\infty)$. Furthermore, from \eqref{fx} and \eqref{gx}, both $f(x)$ and $g(x)$ are polynomials of degree $N-1$. We have:
        \begin{align}
            f(x)&=(1 + u) \sum_{j=0}^{N-2} \left(H_N + \frac{1}{N - 1 - j}\right) u^j\\
            g(x) &= \sum_{j=0}^{N-1}u^j\\
            \Rightarrow \lim_{\theta \rightarrow +\infty} \frac{f(x)}{g(x)} &= H_N + 1 \\
            \Rightarrow \lim_{\theta \rightarrow+\infty}SW(\theta) &= +\infty
        \end{align}
    \end{itemize}

    \item $\mathcal{K} < 0$:
    \begin{itemize}
        \item When $a>1$:
        $SW(\theta)$ has a phase transition, similar to $E_r(\theta)$ on $[\theta_0, +\infty)$, as stated in Theorem 3.2 of \cite{duonghan2021cost}. However, $SW(\theta)$ uses a different threshold value:
        $$
        \beta^*_{SW(\theta)}=-\dfrac{F^*}{\mathcal{K}}
        $$
        \item When $a<1$:
        \begin{itemize}
            \item For $\beta \le \beta^{*}$, $SW(\theta)$ is non--increasing on $[\theta_0, +\infty)$. We have:
    \[
    \max_{\theta \ge \theta_0} SW(\theta)=SW(\theta_0).
    \]
            \item For $\beta > \beta^{*}$, the number of changes of the sign of $\dfrac{dSW(\theta)}{d\theta}$ is at least two for all $N$ and there exists an $N_0$ such that the number of changes is exactly two for $N \le N_0$. As a consequence, for $N \le N_0$, there exist $\theta_1 < \theta_2$ such that, for $\beta > \beta^{*}$, $SW(\theta)$ is decreasing when $\theta < \theta_1$, increasing when $\theta_1 < \theta < \theta_2$ and decreasing when $\theta > \theta_2$. Thus, for $N \le N_0$,
    \[
    \max_{\theta \ge \theta_0}SW(\theta) = \max\{SW(\theta_0), SW(\theta_2)\}.
    \]
        \end{itemize}
        
    \end{itemize}
\end{enumerate}

\subsubsection{Influence of Selection Intensity (\texorpdfstring{$ \beta $}))}

Beside the cost of incentives $\theta$, the intensity of selection $\beta$ also plays a crucial role in deciding the overall stability and long--term optimal structure of the system's Social Welfare ($SW$).

\begin{itemize}
    \item \textbf{Neutral Selection Limit ($\beta \rightarrow 0^+$)}
    
    Under the limit of neutral selection, the incentive value $u$ collapses to 1, causing the optimal incentive to default to zero expenditure.
    We first note that the derivative of $SW$ w.r.t $\theta$ depends directly on $u=e^x=e^{\beta(\theta+\delta)}$ (since $g(x)$ is a polynomial of variable $u$), not $\theta$. We have:
    \begin{align*}
        \lim_{\beta \rightarrow 0^+}u&=\lim_{\beta \rightarrow 0^+}e^{\beta(\theta+\delta)}=1\\
        \intertext{The derivative of the social welfare converges to a constant value, with the same sign as $a-1$, because $P(1) < 0$ and $F(1) < 0$} 
        \lim_{\beta \rightarrow0^+}\frac{dSW(\theta)}{d\theta}&=\frac{N^2P(1)}{2g^2(1)}(a-1)F(1)
    \end{align*}
    This indicates that under the neutral selection limit, $SW$ becomes a strictly monotonic function of $\theta$, or a constant. With $a<1$, since the slope is negative, the maximum social welfare is achieved at the minimum possible incentives: $\max_{\theta}SW(\theta)=SW(0)$.
    
    \item \textbf{Strong Selection Limit ($\beta \rightarrow +\infty$)}
    
    With fixed $\theta$ and approaching deterministic selection, the Dynamics Factor $\Psi(x) = f(x)/g(x)$ reaches its maximum theoretical value (for $\theta > -\delta$):
    \begin{align*}
    \lim_{\beta \rightarrow +\infty}\frac{f(x)}{g(x)}&=H_N +1\\
    \intertext{The expected total social welfare simplifies to a linear function of the incentive cost $\theta$:}
    \lim_{\beta \rightarrow +\infty}SW &= \frac{N^2}{2}\big(H_N+1\big)\big(\delta + N\Delta + (a-1)\theta\big)
    \end{align*}
    This indicates that under deterministic selection, the optimisation problem simplifies: the long--term expected proportion of cooperators stabilizes, and maximizing $SW$ is achieved by simply maximizing the total revenue/efficiency term $(\delta + N\Delta + (a-1)\theta)$. The Dynamics Factor acts only as a final constant multiplier.
\end{itemize}

\subsubsection{Algorithm to Approximate the Optimal Incentive for Maximum Social Welfare}

In this section, we present our approach to approximate the optimal incentive $\theta^*$ that maximizes the Social Welfare function in a Donation Game setting.

Recall that for the expected incentives on reward is modelled as:
$$E_r(\theta) = \frac{N^2 \theta}{2} \frac{\sum_{j=0}^{N-1} \eta_je^{jx}}{\sum_{j = 0}^{N-1} e^{jx}} $$

where $\eta$ is defined as:
\begin{align*}
\eta_0 &= \frac{1}{N-1} + H_{N} \\
\eta_j &= 2H_{N} + \frac{1}{N-j} + \frac{1}{N-j-1} \quad \text{for } 1 \le j \le N-2\\
\eta_{N-1} &= 1 + H_{N}
\end{align*}

with $x = \beta(\theta + \delta)$ and $H_{N} = \sum_{k=1}^{N-1} \frac{1}{k}$ being the well--known harmonic number. (see the supplementary material).

In case of a Donation Game, we can similarly calculate the Social Welfare function as:
$$SW(\theta) = \frac{N^2 (b - c + (a-1) \theta)}{2} \frac{\sum_{j=0}^{N-1} \eta_j e^{jx}}{\sum_{j = 0}^{N-1} e^{jx}}$$

with $x = \beta(a \theta + \delta)$, taking into account the efficiency of the incentive $a$.

We define
\begin{align*}
A(\theta) &:= \frac{N^2}{2} (b - c + (a-1) \theta), \text{ and} \\
B(\theta) &:= \frac{\sum_{j=0}^{N-1} \eta_j e^{jx}}{\sum_{j = 0}^{N-1} e^{jx}}
\end{align*}

Thus, we have $SW(\theta) = A(\theta) B(\theta)$.

Notice that $B(\theta)$ is strictly positive for all $\theta \ge 0$ since all terms in the sums are positive. Therefore, the sign of $SW(\theta)$ depends solely on $A(\theta)$. Furthermore, we are interested in the region which the social welfare is positive, i.e., $SW(\theta) > 0$. This occurs if and only if:
$$A(\theta) > 0 \implies b - c + (a-1) \theta > 0$$

Which corresponds to the half interval $I := [0, \frac{b - c}{1 - a})$ in case $a < 1$ and to the whole set of positive reals in case $a > 1$, which we then notice is trivial since we can just take $\theta \to +\infty$ to maximize social welfare.

Thus, from now on, we focus on the case where $a < 1$ and restrict our analysis to the interval $I$.

Now, consider $\theta \in I$, by taking the natural logarithm of $SW(\theta)$, we have:
\begin{align*}
\log (SW(\theta)) &= \log (A(\theta)) + \log (B(\theta))\\
&= \log (\frac{N^2}{2}) + \log (b - c + (a-1)\theta) + \log \left( \sum_{j=0}^{N-1} \eta_j e^{jx} \right) - \log \left( \sum_{j = 0}^{N-1} e^{jx} \right)
\end{align*}
taking the derivative with respect to $\theta$, we get:
$$\frac{SW'(\theta)}{SW(\theta)} = \frac{a - 1}{b - c + (a-1)\theta} + a \beta \left( \frac{\sum_{j=0}^{N-1} \eta_j j e^{jx}}{\sum_{j = 0}^{N-1} \eta_j e^{jx}} - \frac{\sum_{j=0}^{N-1} j e^{jx}}{\sum_{j = 0}^{N-1} e^{jx}} \right)$$

We wish to find the critical points of $SW(\theta)$, i.e., the values of $\theta$ such that $SW'(\theta) = 0$. Since $SW(\theta) > 0$ for all $\theta \in I$, this is equivalent to finding $\theta$ such that:
\begin{align}
\frac{1 - a}{b - c + (a-1)\theta} = a\beta \left( \frac{\sum_{j=0}^{N-1} \eta_j j e^{jx}}{\sum_{j = 0}^{N-1} \eta_j e^{jx}} - \frac{\sum_{j=0}^{N-1} j e^{jx}}{\sum_{j = 0}^{N-1} e^{jx}} \right) \label{critical points}
\end{align}
We observed that starting from a value $\beta = \beta_0$, the optimal $\theta$ for maximizing social welfare varies little in many cases. This inspired us to approximate the optimal $\theta$ in the general case by searching in a local range starting from the optimal $\theta_{\infty}$ found at $\beta = +\infty$. Now it's our job to find such starting value. We consider 3 cases:

\textbf{Case 1:} If $x_0 = \beta (a \theta_{\infty} + \delta) > 0$, then as $\beta \to +\infty$ the right--hand side of \eqref{critical points} becomes:
$$a\left( \frac{\sum_{j=0}^{N-1} \beta \eta_j j e^{jx}}{\sum_{j = 0}^{N-1} \eta_j e^{jx}} - \frac{\sum_{j=0}^{N-1} \beta j e^{jx}}{\sum_{j = 0}^{N-1} e^{jx}} \right)$$

dividing both numerator and denominator by $e^{(N-1)x}$, the expression becomes:
$$ a\left( \frac{\sum_{j=0}^{N-1} \beta \eta_j j e^{(j - N + 1)x}}{\sum_{j = 0}^{N-1} \eta_j e^{(j - N + 1)x}} - \frac{\sum_{j=0}^{N-1} \beta j e^{(j - N + 1)x}}{\sum_{j = 0}^{N-1} e^{(j - N + 1)x}} \right) = a \frac{U(e^x)}{V(e^x) + \eta_{N-1}} $$

where $U(x) = \sum_{j = -(N-2)}^{-1} u_j x^j$ and $V(x) = \sum_{j = -(N-2)}^{-1} v_j x^j$ are polynomials in $x$ with negative powers, and coefficients that grows linearly with $\beta$ in the case of $U(x)$, and independent of $\beta$ in the case of $V(x)$. Therefore, since  $x$ grows exponentially with $\beta$, as $\beta \to + \infty$ both polynomials vanish, resulting in the final limit being zero, which forces the left--hand side to do so as well. This gives us:
$$ \frac{1 - a}{b - c + (a-1)\theta_{\infty}} = 0$$
which is impossible.

\textbf{Case 2:} If $x_0 < 0$, then a similar argument shows that as $\beta \to +\infty$, the right--hand side of \eqref{critical points} becomes 0 as well, leading to the same contradiction.

\textbf{Case 3:} $x_0 = 0$. This is the only possible case, i.e.,
$$a \theta_{\infty} + \delta = 0 \implies \theta_{\infty} = -\frac{\delta}{a}$$

Notice that for $\theta_{\infty}$ to be in the interval $I$, we need: $ -\frac{\delta}{a} < \frac{b - c}{1 - a}$, which if we substitute in $\delta = - (c + \frac{b}{N - 1})$ gives:
$$\frac{c + \frac{b}{N - 1}}{a} < \frac{b - c}{1 - a}$$
which rearranges to:
$$c < \frac{b(Na - 1)}{N - 1}$$

Therefore if this condition is not satisfied, we start our search from the boundary point $\theta_{start} = \frac{b - c}{1 - a}$. Otherwise, we start from $\theta_{start} = \theta_{\infty} = -\frac{\delta}{a}$.

Below is the pseudocode of our simple local search algorithm (Algorithm \ref{alg:pgg_main}) to find an approximation of the optimal incentive $\theta^*$ for maximizing social welfare.


    

    



\begin{algorithm}[H]
\caption{Restricted Interval Grid Search with Upper Bound Cutoff}
\label{alg:pgg_main}
\KwIn{Parameters $N \in \mathbb{Z}^{+}$ and $a, b, c \in \mathbb{R}$, where $a < 1$ and $b > c$. Search radius $\varepsilon > 0$ and number of steps $S \in \mathbb{Z}^{+}$}
\KwOut{$\theta^{*}$ maximizing SW within valid bounds}

\textbf{Step 1: Compute Bounds}

$\;\;\;\;\;\;\delta \gets -(c + \frac{b}{N-1})$;

$\;\;\;\;\;\;\theta_{limit} \gets \frac{b-c}{1-a}$ \tcp*[r]{Define the hard upper limit}

$\;\;\;\;\;\;\theta_{start} \gets \min \left( \frac{-\delta}{a}, \theta_{limit} \right)$;

\textbf{Step 2: Initialize Grid Search}

$\;\;\;\;\;\;\theta_{best} \gets \text{null}$;

$\;\;\;\;\;\;SW_{best} \gets - \infty$;

$\;\;\;\;\;\;\Delta \theta \gets \frac{2 \varepsilon}{S}$ \tcp*[r]{Calculate step size}

\textbf{Step 3: Evaluate Interval}

\For{$k \gets 0$ \textbf{to} $S$}{
    $\theta_{curr} \gets (\theta_{start} - \varepsilon) + k \cdot \Delta\theta$;

    \If{$\theta_{curr} > \theta_{limit}$}{
        \textbf{break} \tcp*[r]{Terminate search if limit is exceeded}
    }

    $val \gets SW(\theta_{curr})$;

    \If{$val > SW_{best}$}{
        $SW_{best} \gets val$;

        $\theta_{best} \gets \theta_{curr}$;
    }
}

\textbf{Step 4: Return Result}

\Return{$\theta_{best}$};
\end{algorithm}

\subsection{Numerical Simulation for well--mixed populations}
\begin{figure*}[tb]
    \centering
    \includegraphics[width=\linewidth]{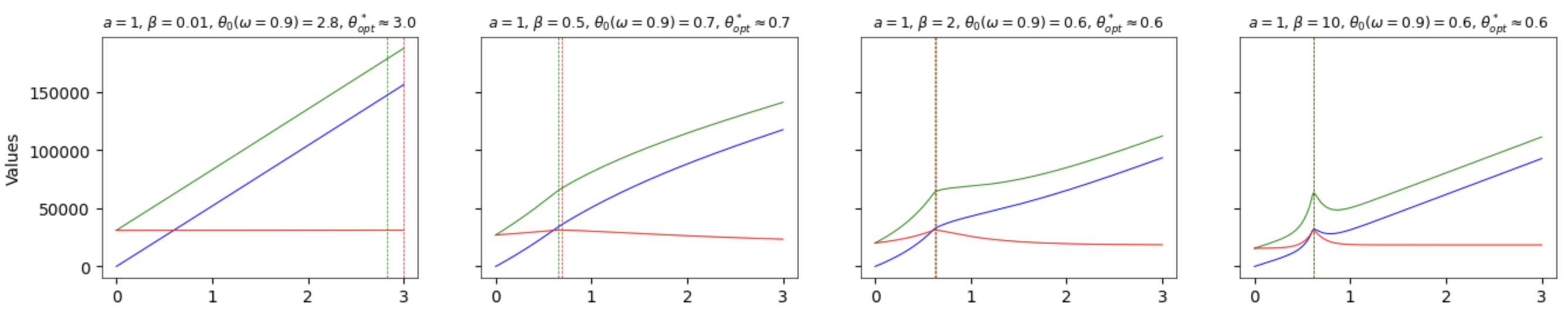}
    \includegraphics[width=\linewidth]{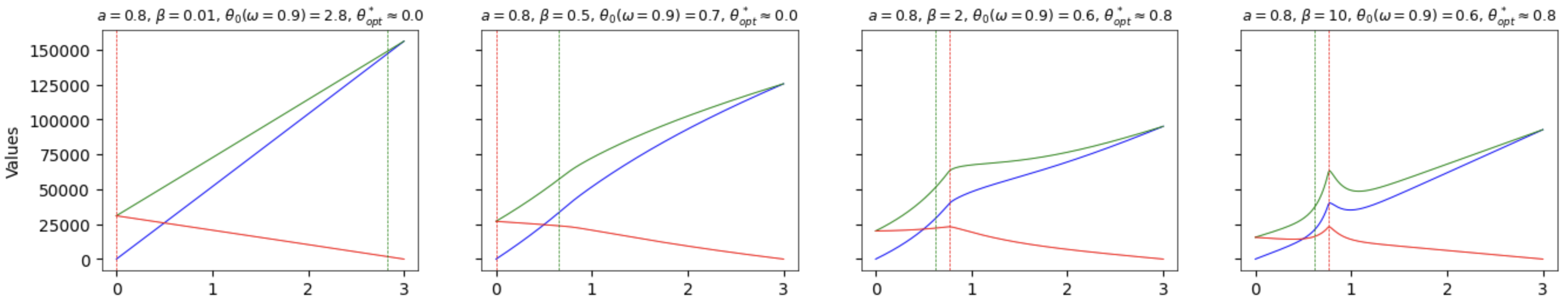}
    \includegraphics[width=\linewidth]{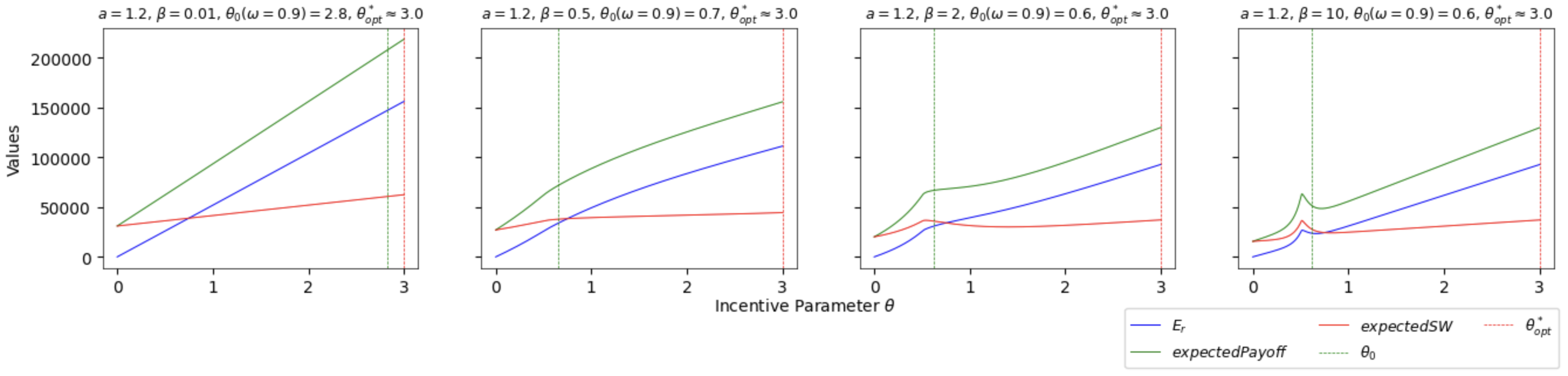}
    \caption{
    The expected total Social Welfare (solid red), the expected total Institution Cost (solid blue) and the expected total Payoff of the population (solid green), w.r.t incentive $\theta$. The red vertical dashed line represents the maximizer for the Social Welfare. The green vertical dashed line denotes the minimum incentive to acquire the cooperation frequency of $90\%$.
    }
    \label{plots1a}
\end{figure*}

Figure \ref{plots1a} plots the expected total Social Welfare (solid red), the expected total Institution Cost (solid blue), and the expected total Payoff of the population (solid green) with respect to the incentive $\theta$. The simulations use a Donation Game with parameters $b=1.2$ and $c=0.6$. This analysis investigates the observed behaviours of the three optimisation objectives—maximizing Cooperation Frequency ($\omega$), minimizing Institution Cost ($E$), and maximizing Social Welfare ($SW$)—across varying values of the efficiency parameter $a$ ($a=1$, $a<1$, and $a>1$) and selection intensity $\beta$. We specifically analyse the alignment and conflict between the solution that maximizes Social Welfare and the optimal solutions for the other two objectives.

\begin{enumerate}
    \item \textbf{Case $a=1$: Zero--Sum Transfer}
    
    As the intensity of selection approaches neutrality ($\beta \rightarrow 0^+$), the expected Social Welfare is observed to converge to a constant value, rendering the objective to maximize $SW$ effectively redundant as $SW$'s value does not change with $\theta$. For sufficiently high values of $\beta$, the Social Welfare exhibits a unique optimum: a local maximizer that also constitutes the global maximizer for $SW$. This $SW$ maximizer is found to coincide with the global maximizer for Cooperation Frequency and the local minimizer of Institution Cost on the interval $[\theta^0, +\infty)$. The critical point $\theta^0$ is defined theoretically as $\theta^0=\log u_0/\beta - \delta$, where $u_0$ is the unique positive root of the polynomial $P(u)$.

    \item \textbf{Case $a<1$: Inefficient Transfer}
    
    Under weak selection ($\beta \rightarrow 0^+$), the expected Social Welfare is shown to decrease monotonically as the incentive $\theta$ increases. Consequently, the maximizer for $SW$ aligns with the optimal solution for minimizing the Institution Cost on the interval $[\theta_0, +\infty)$. For sufficient values of $\beta$, $SW$ develops a unique local maximizer, which is also the global maximizer in the plotted cases. This $SW$ maximizer is observed to slightly diverge from the optimal solution for minimizing the Institution Cost on $[\theta_0, +\infty)$.

    \item \textbf{Case $a>1$: Efficient Transfer}
    
    Regardless of the value of $\beta$, the expected Social Welfare with $a>1$ is observed to increase indefinitely as $\theta \rightarrow +\infty$. This implies that maximizing the incentive $\theta$ always yields the highest Social Welfare for the entire population. This behaviour aligns with the objective of maximizing Cooperation Frequency, as both objectives share the same optimal direction (increasing $\theta$). However, both objectives now conflict with minimizing the Institution Cost, as the lowest cost is achieved by decreasing $\theta$, which is opposite to the optimal direction for Social Welfare and Cooperation Frequency.
\end{enumerate}

The observed behaviours of the expected Social Welfare across all three cases are consistent with the established theoretical analysis conducted on the interval $[\theta^0, +\infty)$. Further analytical exploration of the behaviours on the initial interval $(0,\theta^0]$ is still required.

\subsection{Agent--based simulations for structured populations}

As described in Section \ref{agent-simulation-setup}, we run simulations for a population of size $10000$ in
the lattice graph setup, so at beginning of each round, each agent plays with at most 4 other
neighbouring agents. After each agent gets their payoff result, the external institution steps in and
invest each cooperator an amount $\theta$ corresponding to either the POP or the NEB scheme, thereby
influencing the decision of the agents to choose which behaviour to do in the next game round. To
ensure comparability, all following simulations assume the normalized payoff matrix for Prisoner's
Dilemma game and all other settings the same as \cite{han2018fostering}. A notable addition in the
following results compared to the previous work is the efficiency coefficient $a$ that denotes how
effective the incentivisation is. Unless otherwise specified in the experiments, we set $a=1$, so the social welfare equals the population payoff.

\begin{figure*}[tb]
    \centering
    \includegraphics[width=\linewidth]{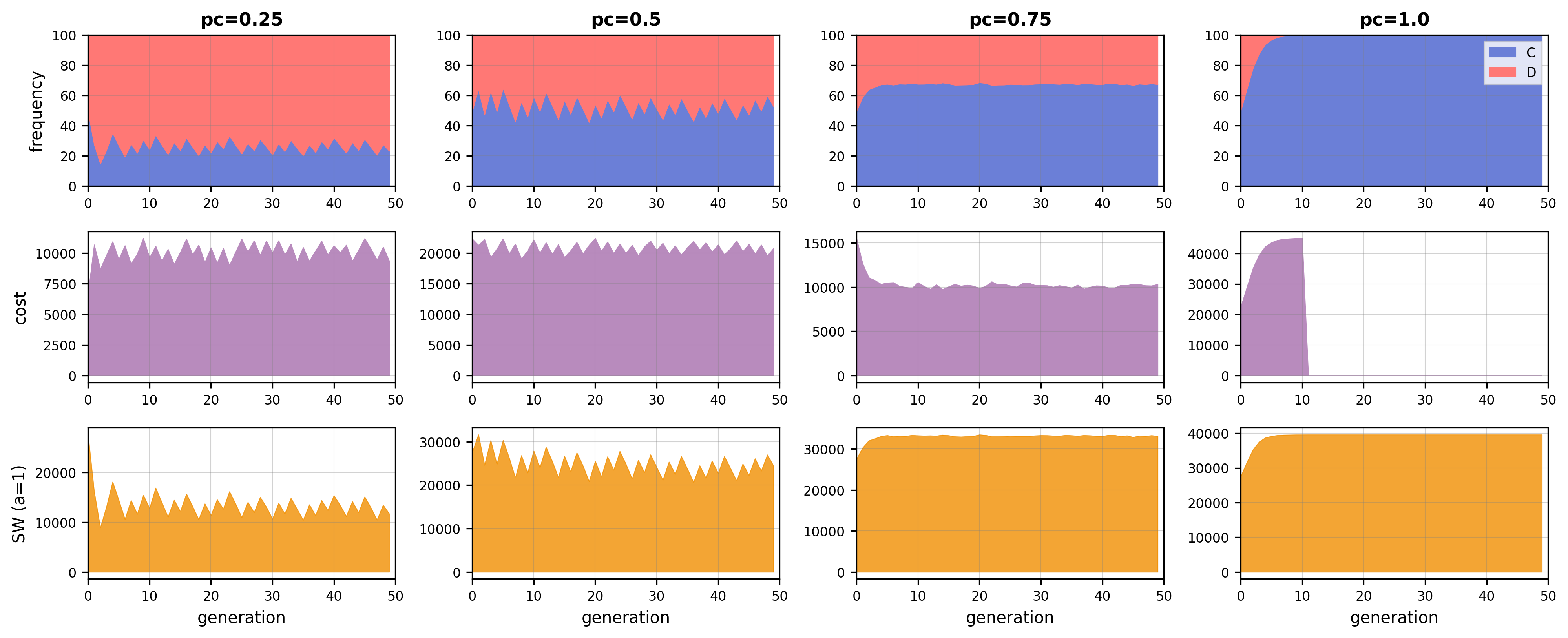}
    \includegraphics[width=\linewidth]{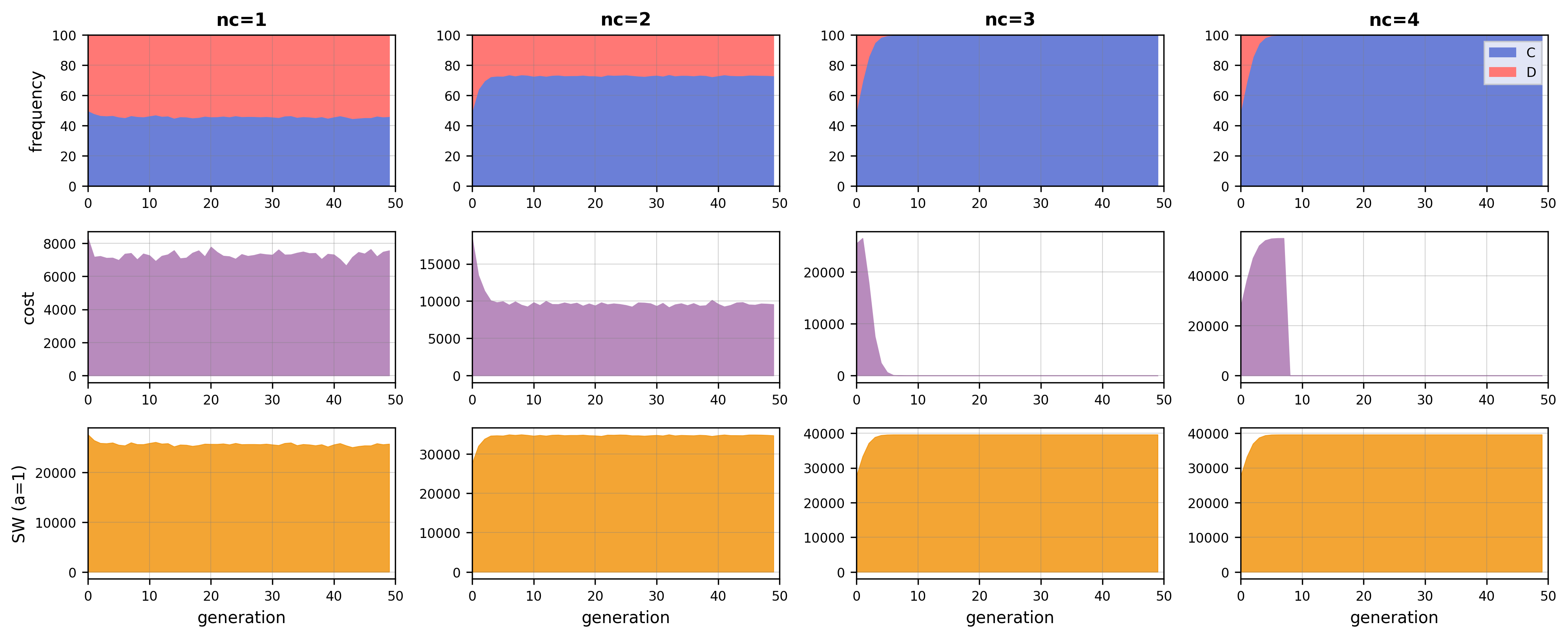}
    \caption{
    Evolution strategies over time (blue for C, red for D), per--generation cost and social welfare.
    (Three top rows) POP for different values of $p_C$, with $\theta = 4.5$.
    (Three bottom rows) NEB for different values of $n_C$, with $\theta = 5.5$.
    We only show 50 generations for the sake of clear presentation. This chart aligns well with the
    findings in \cite{han2018fostering} where POP interventions only work for $p_C$ very close to 1,
    whereas NEB interventions is more cost--efficient with $n_C = 3$ being able to push the
    population to a converged state.
    }
    \label{fig:sim_pop_by_generation}
\end{figure*}

\begin{figure*}[tb]
    \centering
    \includegraphics[width=\linewidth]{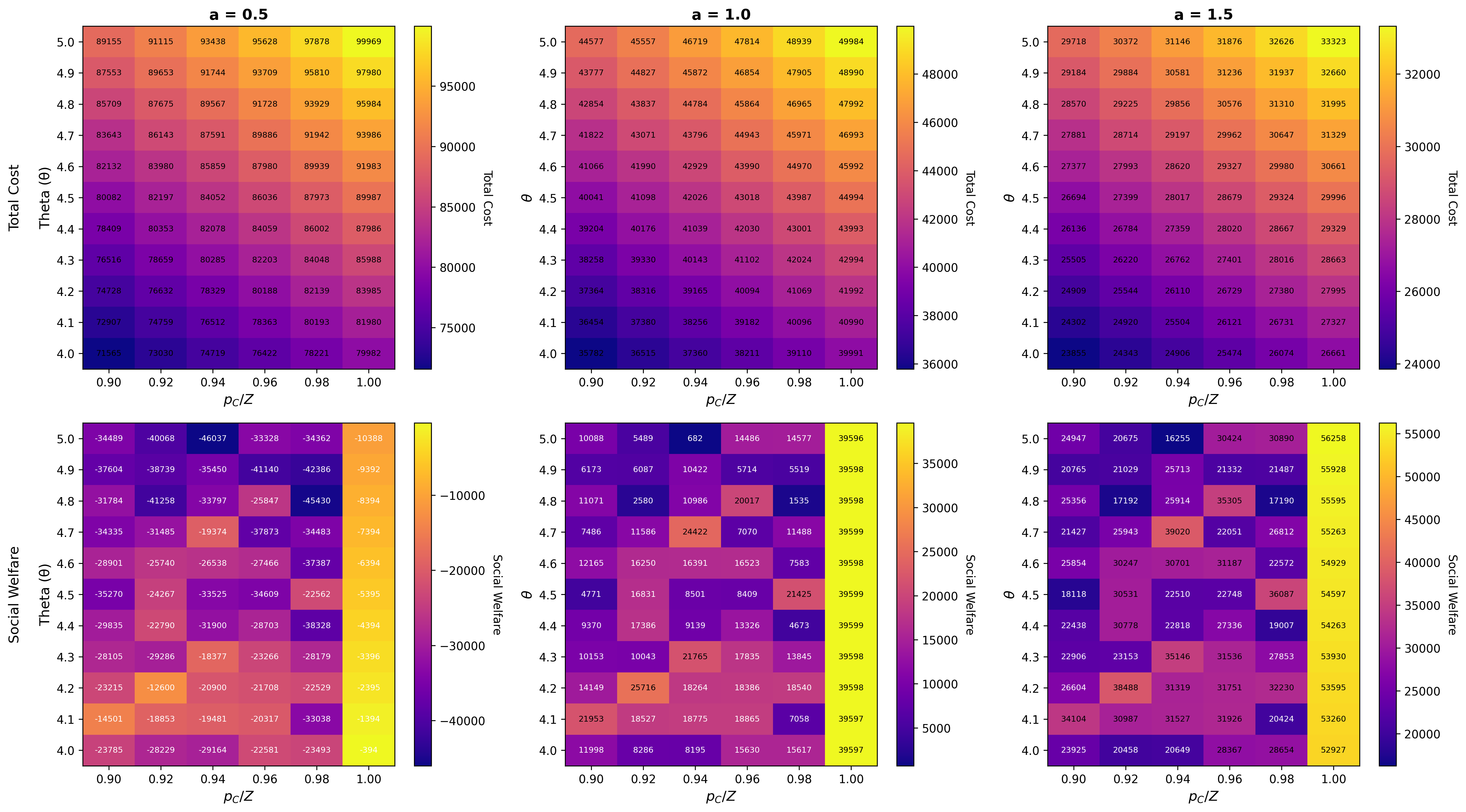}
    \caption{
    POP interference effect on cost and social welfare over multiple efficiency $a$, only
    considering the cases where the cooperation frequency is over 90\%. As POP strategy only pushes
    the whole population to 100\% cooperation at $p_C / Z$ very close to 1, and the experiment
    is deterministic, it is expected that for $a \leq 1$, minimization of cost coincide with
    maximization of social welfare, and for $a > 1$, where the social welfare increases as spending
    increases, minimization of cost would not lead to maximization of social welfare. It is more
    nuanced in the cases where the cooperation frequency is less than 100\%.
    }
    \label{fig:sim_pop_by_sweeping_efficiency}
\end{figure*}

Since $a=1$ means that the social welfare is equal to the population payoff, varying the efficiency
in relation with the different $p_C$ and $n_C$, combined with the different $\theta$ could yield
more insightful information. Figure \ref{fig:sim_pop_by_sweeping_efficiency} presents the POP
intervention method across different $a$, $\theta$ and $p_C$. For the sake of presentation, we only plot
in cases where $p_C / Z \geq 0.9$ which ensures that the population converges around the state where at
least 90\% of the population is cooperative. At $p_C / Z = 1$, the result is predictable where with
$a \leq 1$, as the cost decreases, the social welfare increases, where it is the other way around when 
$a > 1$ (an investment into the population yields better return than the actual spending, or
another way to put it is the perceived cost is higher than the actual cost), in which case as the
cost increases, the social welfare also increases. So for POP, this indicates where minimizing cost
does not result in maximising social welfare.

It becomes more nuanced with $p_C / Z < 1$, where if we pick a value and go from the top down, it
means that the cost will get minimised for roughly the same cooperation level, but the social
welfare demonstrates pockets of values that, for that specific $\theta$, the social welfare is
larger compared to the value that corresponds to the lowest cost, which signifies that the $\theta$
value for maximising the social welfare is different from the $\theta$ value that yields the lowest
cost of intervention. This happens across different $p_C / Z$, so for a relaxed condition where the
population does not fully convert to being cooperative, we can still achieve better social welfare
if we choose a different per--capita cost, thereby ensuring the whole population is better off even
in a suboptimal cooperative social norm.

\begin{figure*}[tb]
    \centering
    \includegraphics[width=\linewidth]{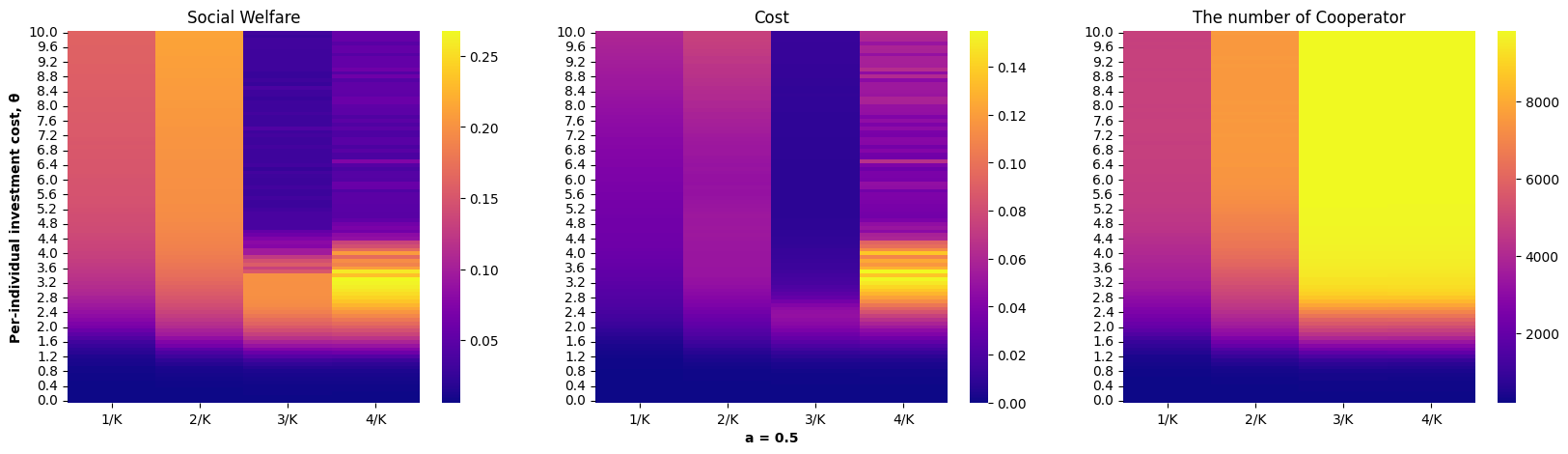}
    \includegraphics[width=\linewidth]{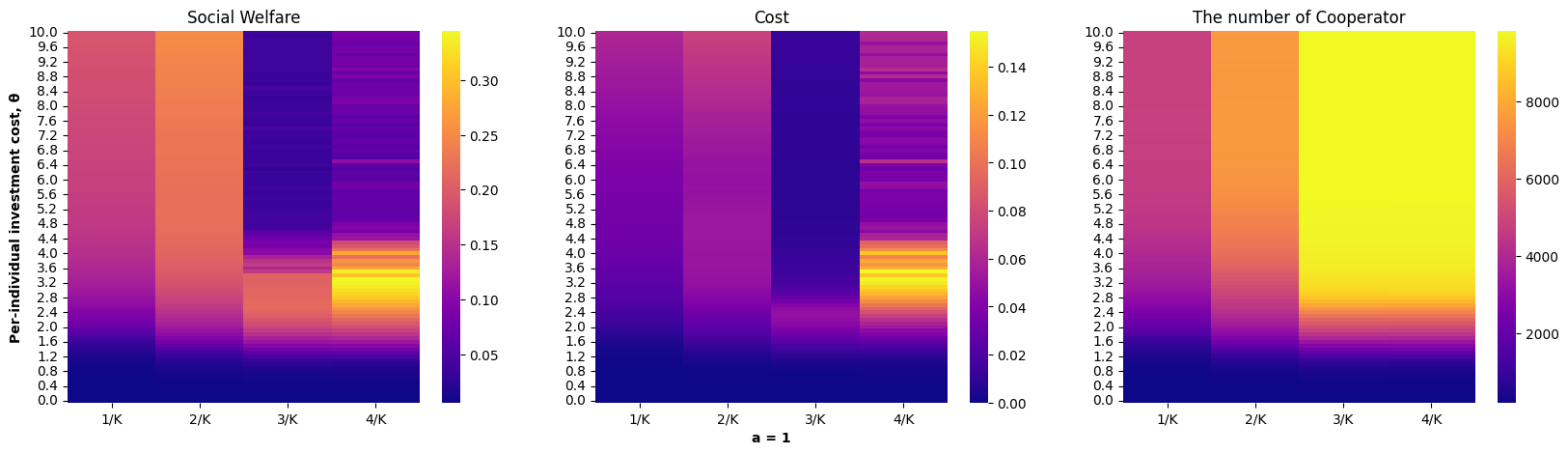}
    \includegraphics[width=\linewidth]{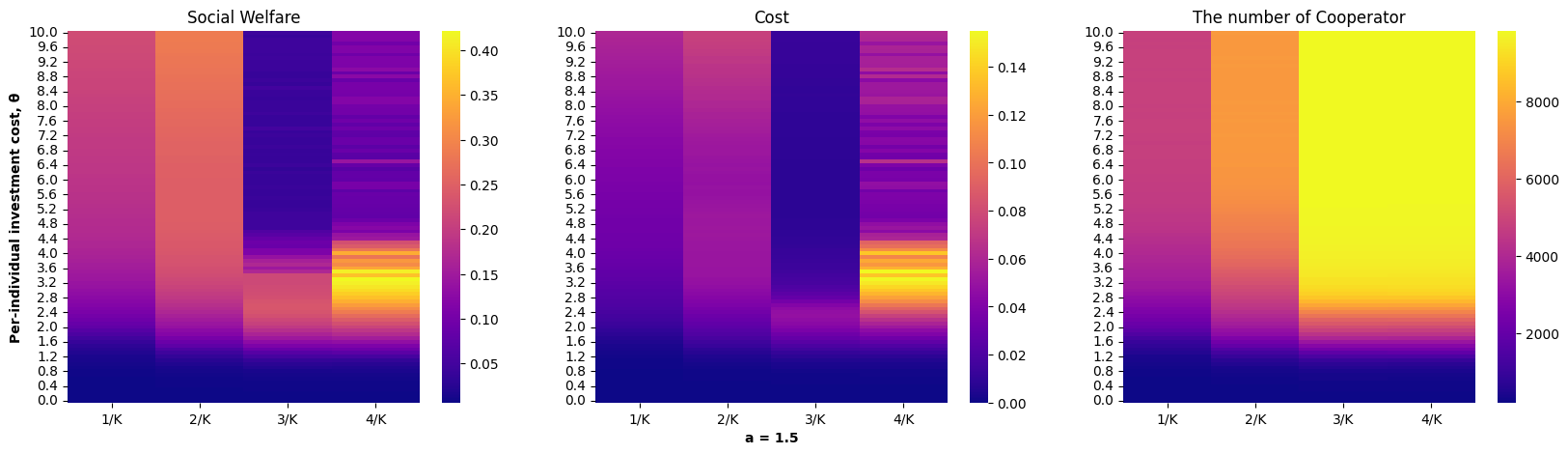}
    \caption{
    The number of cooperators in the neighbourhood trend, the total cost trend, and the total social welfare trend (all are normalized to get the value in range 0 and 1 for clear visualization) for varying per--individual cost of investment $\theta$. The first row is for $n_C = 3$ and the second is $n_C=4$, the left panels are the trend for all $\theta$ and the right ones are the zoom in versions for clearer inspection. }
    \label{fig:heatmap-neb-a}
\end{figure*}

\begin{figure*}[tb]
    \centering
    \includegraphics[width=\linewidth]{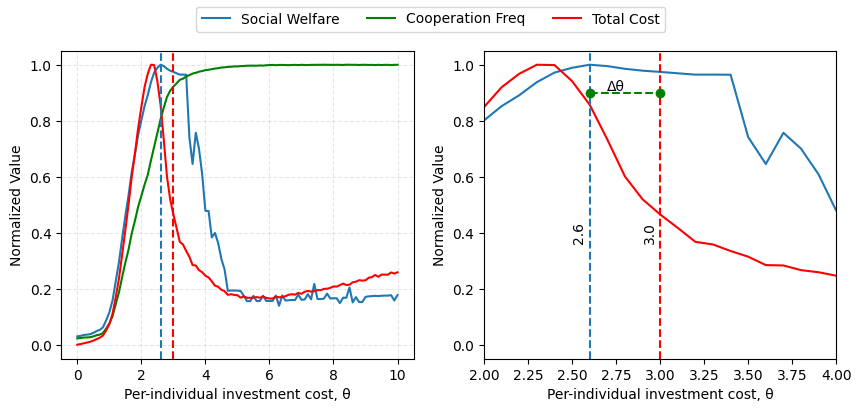}
    \includegraphics[width=\linewidth]{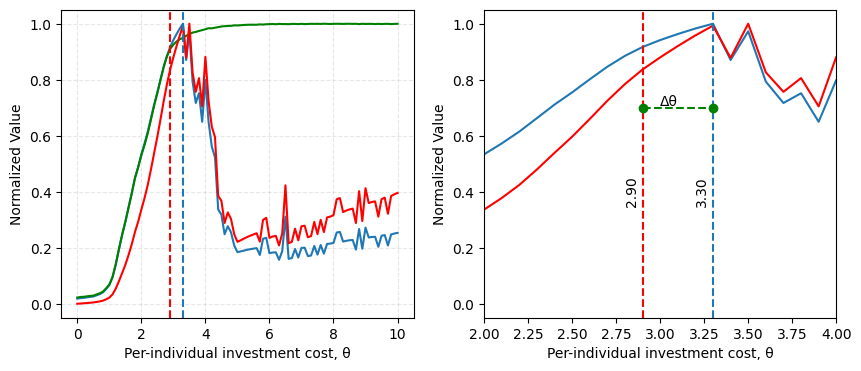}
    \caption{
    The number of cooperators in the neighbourhood trend, the total cost trend, and the total social welfare trend (all are normalized to get the value in range 0 and 1 for clear visualization) for varying per--individual cost of investment $\theta$. The first row is for $n_C = 3$ and the second is $n_C=4$, the left panels are the trend for all $\theta$ and the right ones are the zoom in versions for clearer inspection. The dashed red line locates the $\theta$ for minimal cost and the dashed blue locates the $\theta$ for maximal social welfare. Parameters: $b = 1.8, \ L = 100; \ k = 4$ (node degree)
    }
    \label{fig:different-theta-neb}
\end{figure*}

For NEB intervention strategy, we compare the optimal investment cost $\theta$ for minimizing the total cost and boosting the cooperation frequency in population and the one for maximising social welfare in order to prove that they are not the same. From figure \ref{fig:heatmap-neb-a}, we can observe how NEB obtain the objective in three cases of the cost effective coefficient: $a=0.5, a=1$, and $a=1.5$ on the varying value of $\theta$. We can see that the pattern for NEB to successfully reach the highest frequency of cooperation is similar to what had been explored in \cite{han2018fostering}. The more additional important observation from the first column (the social welfare) is that the social welfare when $n_C=4$ is optimal when the cost is not efficient, this remain the same in all three values of $a$. Moreover, with $n_C=3$, the lower value of $a$, the more value of social welfare added (the orange colour changes from light to dark in the 3/K column when $2.0 < \theta < 3.5$ of the first diagram). At those of $\theta$ values is less cost--efficient.

For clearer observation, as shown in figure \ref{fig:different-theta-neb}, with $n_C=3$, the optimal $\theta$ for cost--efficiency is approximately 2.6 while that for social welfare optimisation is 3.0, the similar pattern can also be seen when $n_C = 4$ (these $\theta$ value are determined when the number of cooperators in the population is greater than 90\% of the total). The gap between two optimal $\theta$ state that when we try to find the efficient value of $\theta$ for concurrently minimizing the cost and boosting the frequency of cooperation, the maximal value of social welfare is not obtained. In addition, at the $\theta$ to reach the maximal value of social welfare, the cost at that point is always higher than the optimal cost in both context of $n_C$. This proves the hypothesis that optimizing both cooperation frequency and cost does not take the social welfare to the efficient state.

\section{Conclusions and Future Work}

\def\infinity{\rotatebox{90}{8}}

In this work, we have approached the optimisation of social welfare in well-mixed and structured populations under external institutional investment from two complementary perspectives: one using mathematical and numerical analysis, and the other employing agent-based simulation. Our goal is to investigate whether optimising intervention costs leads to an optimisation of social welfare.

Both perspectives reach a consensus that the $\theta$ for optimising intervention costs differs from that for optimising social welfare. From the mathematical analysis in well-mixed populations, we have examined the objective function with incentive costs approaching infinity. We have developed an approximation algorithm for the optimal solution in the case of inefficient transfer within the reward mechanism and employed numerical methods to simulate examples illustrating the correlation between the bi-objectives of minimising costs, increasing cooperation, and maximising social welfare. Moreover, we conducted agent-based simulations on a square lattice structured network of the population to observe and analyse the differences in the $\theta$ values needed to achieve the desired configurations under both global and local interference strategies. The simulation  results indicate a gap between the two, thereby supporting our hypothesis regarding the effectiveness of multi-objective approaches on optimal social welfare.

As future work, we will investigate the complete behaviour of the social welfare objective function in $\mathbb{R}^+$ and seek to develop an algorithm that approximates the solution for the case of zero-sum transfers within the reward mechanism. 
Additionally, we plan to conduct simulations on multiplayer games, such as the Public Goods Game, to examine the patterns of various criteria (total cost, social welfare) and gain deeper insights into the effects of network structure on these objectives.

\section*{Acknowledgements}
This work was produced during the workshop ``HUMAN BEHAVIOUR MODELLING AND AI AGENTS USING GAME THEORY'', organised at HCMUT--VNUHCM.
T.A.H. acknowledges travel support from the HCMUT--VNUHCM (Adjunct Professorship scheme HCMUT--VNUHCM). Z.S. and TAH are supported by EPSRC (grant EP/Y00857X/1).

\section*{Competing interest} Authors declare that they have no conflict of interest.


\clearpage
 
\bibliographystyle{unsrt}
\bibliography{mybib}

\newpage

\setcounter{figure}{0}
\setcounter{equation}{0}

\renewcommand*{\thefigure}{A\arabic{figure}}
\renewcommand*{\theequation}{A\arabic{equation}}


\end{document}